\documentstyle[12pt,aaspp4]{article}

\begin{document}

\title{RXTE Spectral Observations of the 1996--97 Outburst of the Microquasar 
GRO~J1655--40}
\author{Gregory J. Sobczak}
\affil{Astronomy Dept., Harvard University, 60 Garden St., Cambridge, MA
02138; gsobczak@cfa.harvard.edu}
\author{Jeffrey E. McClintock}
\affil{Harvard-Smithsonian Center for Astrophysics, 60 Garden St., Cambridge, MA
02138; jem@cfa.harvard.edu}
\author{Ronald A. Remillard}
\affil{Center for Space Research, MIT, Cambridge, MA 02139; rr@space.mit.edu}
\author{Charles D. Bailyn}
\affil{Department of Astronomy, Yale University, P. O. Box 208101, New Haven, CT
06520; bailyn@astro.yale.edu}
\author{Jerome A. Orosz}
\affil{Dept. of Astronomy and Astrophysics, The Pennsylvania State University, 525 Davey
Laboratory, University Park, PA 16802; orosz@astro.psu.edu}

\begin{abstract}

Excellent coverage of the entire 16-month 1996--97 outburst cycle of GRO~J1655--40 was
provided by the {\it Rossi X-ray Timing Explorer}.  We present a full spectral
analysis of these data, which includes 52 Proportional Counter Array spectra from
2.5--20~keV and High Energy X-ray Timing Experiment spectra above 20~keV.  We also
include a nearly continuous All-Sky Monitor light curve with several intensity
measurements per day.  The data are interpreted in the context of the multicolor
blackbody disk/power-law model.  The source exhibits two principal outburst states
which we associate with the very high and the high/soft states.  During the very high
state, the spectrum is often dominated by a power-law component with photon index
($\Gamma$) $\sim$~2.3--2.7.  The source exhibits intense hard flares on time scales of
hours to days which are correlated with changes in both the fitted temperature and
radius of the inner accretion disk.  During the high/soft state, the spectrum is
dominated by the soft thermal emission from the accretion disk with spectral
parameters that suggest approximately constant inner disk radius and temperature.  The
power-law component is relatively weak with $\Gamma \sim$~2--3.  During the last few
observations, the source undergoes a transition to the low/hard state.  

We find that a tight relationship exists between the observed inner radius of the disk
and the flux in the power-law component.  During intense hard flares, the inner disk
radius is observed to decrease by as much as a factor of three on a time scale of
days.  The apparent decrease of the inner disk radius observed during the flares may
be due to the failure of the multicolor disk model caused by a steepening of the
radial temperature profile in the disk coupled with increased spectral hardening and
not physical changes of the inner disk radius.  The distortion of the inner disk
spectrum by the power-law flares indicates that the physical mechanism responsible for
producing the power-law emission is linked to the inner disk region.  

Assuming that our spectral model is valid during periods of weak power-law emission,
our most likely value for the inner disk radius implies $a_* < 0.7$.  Such a low value
for the black hole angular momentum is inconsistent with the relativistic frame
dragging and the `diskoseismic' models as interpretations for the 300~Hz X-ray QPO seen
during some of these RXTE observations.

\end{abstract}


\section{Introduction}

The X-ray Nova GRO~J1655--40 was discovered 1994 July 27 (UT) by the Burst and
Transient Source Experiment (BATSE) onboard the {\it Compton Gamma Ray Observatory}
(Zhang et al. 1994).  The optical counterpart was discovered soon thereafter by Bailyn
et al. (1995).  It was subsequently found that GRO~J1655--40 contained a black hole
primary with mass $7.02\pm0.22~M_{\odot}$ (Orosz \& Bailyn 1997; van der Hooft et al.
1998), the most accurately measured mass for any black hole candidate.  GRO~J1655--40
is one of only eight Black Hole X-ray Novae (BHXN) with a dynamically determined
primary mass that exceeds 3~$M_{\odot}$, the maximum mass of a neutron star (Kalogera
\& Baym 1996; McClintock 1998).  GRO~J1655--40 is one of a few Galactic X-ray sources
known to produce superluminal radio jets (Tingay et al. 1995; Hjellming \& Rupen
1995).  The others are GRS~1915+105, which is also suspected of being a black hole
(Mirabel \& Rodriguez 1994,1998), and possibly Cyg X-3 (Mioduszewski et al. 1997;
Newell et al. 1998).  These sources, and a few other Galactic X-ray sources that
exhibit double-lobed radio structures (e.g. 1E1740.7--2942; see Smith et al. 1997 and
references therein), are known collectively as the `microquasars', since they have
properties analogous to radio-loud active galactic nuclei.  GRO~J1655--40 is also of
intense current interest because it has been proposed to be a rapidly rotating Kerr
black hole based on its spectral characteristics (Zhang, Cui, \& Chen 1997a) and on
the frame dragging model for the observed 300~Hz QPO (Cui, Zhang, \& Chen 1998).  

The X-ray behavior of BHXN can be described in terms of five distinct, canonical
spectral states characterized by the presence or absence of a soft blackbody component
at $\sim$~1~keV and the properties of a power-law component at higher energies above
$\sim$~10~keV.  In order of increasing luminosity, these states are the {\it
quiescient/off}, {\it low/hard}, {\it intermediate}, {\it high/soft}, and {\it very
high} states.  The {\it quiescient/off} state is characterized by an X-ray luminosity
several orders of magnitude lower than the other states, and a power-law spectrum with
a photon index $\sim 2$ (Narayan, Barret \& McClintock 1997).  The {\it low/hard}
state consists of a power-law component with photon index $\Gamma \sim 1.4$--1.9 with
an exponential cutoff $\sim$~100~keV and a weak thermal component.  Sources in the
{\it low/hard} state exhibit strong variability at frequencies $\lesssim10$~Hz (van
der Klis 1995).  The {\it high/soft} state is dominated by a soft $\sim$~1~keV
blackbody component due to a hot accretion disk (Tanaka \& Shibazaki 1996) and a
power-law with $\Gamma \sim 2.2$--2.7.  The {\it high/soft} state has an X-ray
luminosity $L_x \sim 0.2$--0.3~$L_{Edd}$ compared to $L_x\lesssim0.1~L_{Edd}$ for the
{\it low/hard} state (Nowak 1995).  Sources in the {\it high/soft} state exhibit
little temporal variability (van der Klis 1995).  The {\it intermediate} state, as
evidenced by its name, has properties in common with both the {\it low/hard} and {\it
high/soft} spectral states.  The intermediate state has been observed as a distinct
spectral state in Nova Muscae 1991 (Ebisawa et al. 1994), Cyg~X-1 (Belloni et al. 
1996), and GX~339$-$4 (Mendez \& van der Klis 1997).  The {\it very high} state
spectrum has a dominant power-law component with photon index $\sim 2.5$ and exhibits
strong variability and high X-ray luminosity, $L_x \sim L_{Edd}$ (Ebisawa et al. 
1994).  Sources in the {\it very high} state exhibit quasi-periodic oscillations
(QPOs) near 3--10~Hz and at frequencies as high as 300~Hz (van der Klis 1995;
Remillard et al.  1998).  Thermal emission from the disk remains visible in the {\it
very high} state.

In this paper we present the spectral results for 52 pointed observations which cover
the entire 16-month 1996--97 outburst of GRO~J1655--40 and employed all the
instruments aboard the {\it Rossi X-ray Timing Explorer} (RXTE): namely, the All-Sky
Monitor (ASM), the Proportional Counter Array (PCA), and the High Energy X-ray Timing
Experiment (HEXTE).  Our results consist of ASM lightcurves, 52 detailed PCA spectra
from 2.5--20~keV, and HEXTE spectra above 20 keV when the hard X-ray count rate was
sufficient.  The spectra were fit to a model including interstellar absorption,
multicolor blackbody disk, and power-law components.  A timing study based on these
same RXTE observations of GRO~J1655--40 is presented in a companion paper (Remillard
et al. 1998).  

The observations are discussed in \S~2, and the results of the spectral fitting are
presented in \S~3.  Section~4 highlights the correlations between the disk and
power-law components with a discussion of the limitations of the multicolor disk model
and the interpretation of the fitted inner disk radius.  Section~5 contains a
discussion of the Zhang et al. (1997a) corrections to the observed spectral parameters
due to general relativistic effects and an estimate of the black hole angular
momentum.  The relation between the observed spectral parameters and QPOs is discussed
in \S~6 followed by a brief summary of our results.  

The units used in this paper are $r=R/r_g$ where $r_g=GM/c^2$, $m=M/M_{\odot}$,
$l=L/L_{Edd}$ where $L_{Edd}=1.25\times10^{38}~m$~erg~s$^{-1}$ is the Eddington
luminosity, and $\dot{m}=\dot{M}/\dot{M}_{Edd}$, where
$\dot{M}_{Edd}=L_{Edd}/(0.1 c^2)=1.39 \times 10^{18} m$ g s$^{-1}$ is the Eddington
accretion rate, assuming an accretion efficiency of 10\%.

\section{Observations \& Reductions}

We present observations covering the entire 16 month 1996--97 outburst of
GRO~J1655--40 obtained using the ASM, PCA, and HEXTE instruments onboard RXTE.  The
ASM has three energy channels corresponding to 1.5--3~keV, 3--5~keV, and 5--12~keV
(Levine et al. 1996).  The PCA data were taken in the ``Standard~2'' format which
corresponds to 129 energy channels from 0--100~keV.  The PCA contains five individual
proportional counter units (PCUs 0--4).  The response matrix for each PCU was obtained
from the 1997 October~2 distribution of response files from Keith Jahoda's ftp site on
lheaftp.gsfc.nasa.gov.  The application of more recent response files (1998 January)
to a few of our observations shows that the fit parameters differ by less than 5\%.  

The spectrum from each PCU was fit individually over the energy range 2.5--20~keV,
including a systematic error in the count rates of 1.5\%.  The good energy range was
decided by trial-and-error fitting of archival observations of the Crab Nebula (1997
July~26 \& 1997 March~22) and the data set presented here.  The lower limit of 2.5~keV
was used because of uncertainty in the response at lower energies, and the upper limit
of 20~keV was used because there are systematic problems (at the level of a few
percent) in the response matrices and background subtraction near 25~keV.  The PCA
spectra were background subtracted using the standard background model for bright
sources, which includes corrections for the instantaneous particle flux, activation,
and the cosmic X-ray background.  We found that PCUs~2 \& 3 yielded reduced
chi-squared ($\chi^2_{\nu}$) values consistently higher than PCUs~0, 1, \& 4 for fits
to archival observations of the Crab Nebula.  In addition, we found that PCU~4 gave
consistently lower count rates ($\sim$~3\%) than PCUs 0 \& 1 at low energies
($\lesssim~6$~keV).  Consequently, only PCUs 0 \& 1 were used for the spectral fitting
reported here and both PCUs were fit simultaneously using XSPEC.  

The standard HEXTE reduction software was used for the extraction of the HEXTE data. 
We used the HEXTE response matrices released 1997 March 20.  Only the data above 20
keV was used because of uncertainty in the response at lower energies.  The HEXTE
modules were alternatingly pointed every 32~s at source and background positions,
allowing background subtraction with high sensitivity to time variations in the
particle flux at different positions in the spacecraft orbit.  The HEXTE normalization
is allowed to float independent of the PCA normalization since there is a $\sim20$\%
systematic offset between the two instruments in the normalization for the Crab
nebula, part of which is due to uncertainties in the deadtime for the HEXTE
instrument.  All of the normalizations reported here were obtained from the PCA
data.

\section{Analysis \& Results}

The total ASM lightcurve (2--12~keV) and the ASM spectral hardness ratio, HR2, are
plotted in Figures~1a \& b.  The origin of time was chosen such that day zero
corresponds to the initial rise in the X-ray intensity in the ASM: day 0 = MJD~50198 =
1996 April~25~(UT) (Orosz et al. 1997) (MJD~=~JD~--~2,400,000.5).  The lightcurve
exhibits erratic flaring from approximately day 27 to 147 and again from day 175 to
day $\sim$~200.  The flaring ceases shortly before the gap in the data around day 220,
which is when the source moved into the solar exclusion zone (SEZ).  The
(5--12~keV)/(3--5~keV) ratio is roughly constant following passage through the SEZ,
except for a dip followed by a sharp hardening of the spectrum near day 470.  After
day 470, the PCA data show (see below) that the source enters the low/hard state; it
is then too faint to detect with the ASM.  The outburst of GRO~J1655--40 evolves
through a succession of three of the canonical spectral states (very high, high/soft,
and low/hard), as indicated in Figure~1a.  The states are defined by the timing
behavior and spectral characteristics of each state.  Sample spectra from each of
these states are plotted in Figure~2, along with the best fit models for each
spectrum.  Mendez, Belloni, \& van der Klis (1998) identified three states during the
decay of the outburst using some the of same observations presented here.  

The PCA/HEXTE spectral data were fit using XSPEC to a model of interstellar absorption
using the Wisconsin cross-sections (Morrison \& McCammon 1983) plus a multicolor
blackbody accretion disk (Mitsuda et al. 1984; Makishima et al. 1986) plus a power-law
component.  In addition, a smeared Fe aborption edge or a Compton reflection component
were applied in particular cases, as described below.  The hydrogen column density was
fixed at $0.89\times10^{22}$ atoms per cm$^{-2}$ (Zhang et al. 1997b).  

The multicolor disk $+$ power-law model is widely used and well established (Tanaka \&
Lewin 1995, and references therein; Ebisawa et al. 1994).  Some assumptions and
limitations of the model are discussed in \S~4 and figure prominently in our
interpretation of the observed changes in the inner disk radius.  We attempted to use
several alternative models to fit the observed spectra.  Combinations of disk
blackbody, power-law, thermal bremsstrahlung, and comptonization models resulted
either in values of $\chi_{\nu}^2$ several times larger than the multicolor disk $+$
power-law model or gave unphysical values for the fit parameters.  

The four principle quantities returned from the fits were the power-law photon index
($\Gamma$), the power-law normalization ($K$) in units of photons s$^{-1}$ cm$^{-2}$
keV$^{-1}$ at 1~keV, the color temperature of the inner accretion disk ($T_{col}$) in
keV, and the multicolor disk normalization parameter: 
\begin{equation} 
\left( \frac{R_{col}}{km} \right)^2 / \left(\frac{D}{10~kpc} \right)^2 \cos{\theta}, 
\end{equation} 
where $R_{col}$ is the inner disk radius in kilometers derived from the color
temperature, $D$ is the distance to the source in kiloparsecs, and $\theta$ is the
inclination angle of the system.  The mass, inclination angle, and distance of
GRO~J1655$-$40 are well determined: $M = 7.02\pm0.22~M_{\odot}$,
$\theta=69\fdg5\pm0\fdg1$ (Orosz \& Bailyn 1997), and $D=3.2\pm0.2$~kpc (Hjellming \&
Rupen 1995).  Substituting these values into equation (1) allows us to solve for
$R_{col}$ in units of $r_{col}=R_{col}/r_g$ where $r_g=G M/c^2=10.4$~km.  

$T_{col}$ \& $r_{col}$ (usually referred to in the literature as $T_{in}$ \&
$r_{in}$) are often taken to be the temperature and radius of the inner accretion
disk.  However, corrections for spectral hardening must be made to the observed
spectral parameters ($T_{col}$ \& $r_{col}$) to determine the effective temperature
and radius of the inner disk ($T_{eff}$ \& $r_{eff}$).  These corrections compensate
for the fact that electron scattering dominates over absorption as a source of opacity
in the disk and affects the inner disk spectrum through Comptonization of the emergent
spectrum (Shakura \& Sunyaev 1973).  Such a Comptonized spectrum may be approximated
by a diluted blackbody (Ebisawa et al. 1994): 
\begin{equation} 
I \left( E \right) = \left( \frac{1}{f^4} \right) B \left( T_{col}, E \right), 
\end{equation} 
where $B(T_{col},E)$ is the Planck function, $T_{col}$ is the color temperature, and
$f$ is the color correction or spectral hardening factor.  In this diluted blackbody
spectrum, the spectral shape is the same as a blackbody with temperature
$T_{col}=fT_{eff}$, where $T_{eff}$ is the effective temperature (also called
$T_{peak}$ or $T_{max}$), and the normalization is smaller by the factor $1/f^4$. 
Since the normalization fit by the multicolor disk model is proportional to
$r_{col}^2$ (see Eq. (1)), the actual radius should be $f^2$ times larger than
obtained from fitting the multicolor disk model.  

Shimura \& Takahara (1995) have shown, using a numerical simulation which
self-consistently solves for the vertical structure and radiative transfer in the
accretion disk, that the spectral hardening factor
$f=(T_{col}/T_{eff})$ can be approximated by a constant $f=1.7\pm0.2$ for
$\alpha_v\sim0.1$, $1.4\leq~m~\leq 10$, and $0.1\leq~\dot{m}~\leq~10$, where
$\alpha_v$ is the viscosity parameter.  
Therefore, the fit values of $T_{col}$ and $r_{col}$ can be corrected approximately 
for spectral hardening by using the following formulas:
\begin{equation}
T_{eff} = \frac{T_{col}}{f} = \frac{T_{col}}{1.7},
\end{equation}
\begin{equation}
r_{eff} = f^2 r_{col} = 2.9 r_{col},
\end{equation}
where $T_{eff}$ and $r_{eff}$ are the effective temperature and radius of the inner
accretion disk.  The model parameters, corrected for spectral hardening, are listed in
Table~1 and presented in Figures~3a--d.

\subsection{Very High State}

In Figures~1a \& 3d, we see that between days 27 and 147, and days 175 and 222, there
are sudden flares during which the power-law normalization increases by a factor of
ten in only a few days.  These flares in the hard component are accompanied by an
increase in $T_{eff}$ from $\sim$~0.76~keV to 1.13 keV (Fig.~3a) and a decrease in
$r_{eff}$ from $\sim$~5.5 to $1.6$ (Fig.~3b).  There is also an abrupt increase in
$\Gamma$ from $\sim$~2.3 to 2.7 (Fig.~3c).  We identify this flaring behavior with the
very high state because of the strong power-law component with $\Gamma \sim 2.5$ and
the presence of strong QPOs from 7.3--22.4~Hz (Remillard et al. 1998).  Figure~2a
shows an example of the spectrum during the flares.  Between days 147 and 175 the
power-law (hard) flares cease and the photon index decreases to 2.0--2.2.  During this
time, the effective temperature and radius of the inner disk become steady at $T_{eff}
\sim 0.75$ keV and $r_{eff} \sim 5.5$; Figure~2b shows a representative spectrum. 
Nevertheless, we also identify these observations with the very high state because of
the relatively strong power-law component and the presence of QPOs from 7.3--22.4~Hz,
(Remillard et al.  1998).  The only exceptions are days 154 \& 161 at which time the
source does not exhibit QPOs.  These two observations resemble the high/soft state
discussed in \S~3.2.  

For four observations during the hard flares (days 98, 126, and 191~A \& B), the
fits were significantly improved by adding a Compton reflection component. 
The Compton reflection component was calculated using the `pexriv' model in XSPEC
version~10 (Magdziarz \& Zdziarski 1995).  The resulting values of the disk blackbody
and power-law parameters did not change significantly, and were independent of the
temperature of the reflecting medium.  A 6.4~keV iron $K\alpha$ emission line is
frequently included with the Compton reflection component in such calculations;
however, there was no evidence for 6.4~keV iron $K\alpha$ emission in our spectra. 
The Compton reflection parameters returned from `pexriv' for the four relevent
observations are listed in Table~2.

\subsection{High/Soft and Low/Hard States}

When the source became observable again after passage through the solar exclusion
zone, the power-law normalization had decreased by an order of magnitude relative to
the peak values and was typically $\lesssim~5$~photons s$^{-1}$ cm$^{-2}$ keV$^{-1}$
at 1 keV (Fig.~3d).  The dramatic hard flares had ceased and the effective temperature
and radius of the inner disk had settled down to $\sim$~0.7~keV and 6.5~$r_g$,
respectively (Fig.~3a~\&~b).  The photon index varied significantly from $\sim$~2 to 3
over only a few days (Fig.~3c) and the source exhibited little temporal variability
(Remillard et al. 1998).  We identify this period with the high/soft state.  A sample
high/soft state spectrum is shown in Figure~2c.  

Our initial fits for the high/soft state data were poor ($\chi^2_{\nu}\sim$~1--4);
however, they were significantly improved by the addition of a smeared Fe absorption
edge at 8.0~keV (above the neutral Fe K edge at 7.1~keV) (Ebisawa et al. 1994). 
Figures~4a--c show the effect of the absorption edge on the ratio data/model for a
representative high/soft state spectrum.  From Figures~4a--c, it is apparent that the
addition of an Fe absorption edge significantly improves the $\chi^2_{\nu}$ from 1.44
to 1.04 for an edge at 8.0~keV.  The best fit edge energy varied from 7 to 9~keV for
different spectra.  For consistency the edge was fixed at 8.0~keV with a width of
7~keV and only the Fe optical depth was allowed to float.  The presence of an Fe
absorption feature in BHXN spectra is well established from observations of Cyg~X-1,
GS~2023+338, LMC~X-1, GS~2000+25 (Inoue 1991), and Novae Muscae 1991 (Ebisawa et al.
1994).  

The disk temperature during the high/soft state is approximately constant from
days 307 to 364, then there is a slight increase of $T_{eff}$ from 0.67~keV to 0.71~keV
over 28 days from days 364 to 398 (Fig.~3a).  After this slight rise, $T_{eff}$
decreases from 0.71~keV to 0.60~keV over the next 47 days (Fig.~3a) and $r_{eff}$
increases from 6.3 to 7.1 (Fig.~3b).  

On day 455, the disk temperature begins to drop rapidly and decreases from 0.57~keV to
0.26~keV over the final 32 days.  By day 476 the multicolor blackbody disk appears
only as a soft excess below 5~keV.  The spectra at this time are dominated by the
power-law component with $\Gamma \sim$~1.7--2.0 and the source exhibited increased
variability and low frequency QPOs (Remillard et al. 1998).  These properties are
characteristic of the low/hard state.  A representative spectrum is presented in
Figure~2e.  In the last three observations, the inner disk radius appears to decrease,
which is probably due to the failure of the spectral hardening correction for
$\dot{m}\lesssim0.1$ (Shimura \& Takahara 1995) after day 465.  Similar behavior was
observed in Nova Muscae 1991 (Ebisawa et al. 1994) during the soft-to-hard transition
at late times and is attributed to the same cause.  The apparent decrease in $r_{eff}$
after day 465, just prior to turnoff, can be explained by an increase in the spectral
hardening factor as the inner disk becomes optically thin, rather than a decrease in
the inner disk radius (Ebisawa et al. 1994).  

The values of $\Gamma$ and $K$ for days 14--17 are off the scale used in Figure~3, but
are given in Table~1.  In this initial group of observations, the spectrum below
$\sim$~10~keV is dominated by a soft thermal component with $T_{eff} \sim 0.7$ keV and
by a steep quasi-power-law component with $\Gamma \sim 6$ at higher energies.  Since
the power-law contributes so little flux above 5~keV, compared to the thermal
component, we identify these observations with the high/soft state.  A sample spectrum
from this period is shown in Figure~2d.  The quality of the 2.5--20~keV spectral fits
was poor ($\chi^2_{\nu}\sim$~5--8) during this time.  {\it These spectra resisted all
attempts to apply the multicolor disk plus power-law model, including modifications
for variable $N_H$, emission lines, absorption edges, Compton reflection, thermal
comptonization, and an exponentially cutoff power-law.}  The same difficulty was
encountered for six other observations during the high/soft state.  As a result, we
were forced to fit these spectra using interstellar absorption plus a multicolor disk
over the restricted range 2.5--10~keV, and then fix these parameters when fitting the
high/soft energy component with a power-law from 15--20~keV.  This approach gave
satisfactory values of $\chi^2_{\nu}$ (see Table~1).  Figure~2d shows a sample
spectrum along with the best fit multicolor disk model for these poorly-fit high/soft
state spectra.  In Figures~3a--d, the spectra which are not well fit by the multicolor
disk blackbody plus power-law model above 10~keV are indicated by open circles; in
Table~1 the corresponding entries are flagged by footnotes f and g.  Note that the
effective temperatures and inner disk radii of these cases are consistent with those
of the `normal' high/soft state spectra (Fig.~3a~\&~b).  

The Advection Dominated Accretion Flow (ADAF) model (Narayan \& Yi 1994) has been used
successfully to describe the soft-to-hard transition observed in Nova Muscae 1991
(Esin et al. 1997a) and Cyg X-1 (Esin et al. 1997b).  For Nova Muscae 1991, Esin et
al. (1997a) predicted that the photon index should have risen sharply to $\sim$~4 or 5
in the high/soft state prior to the soft-to-hard transition (see their Fig.~12d),
during the time when Ebisawa et al. were unable to fit the power-law component and
therefore fixed the photon index (see Ebisawa et al. 1994).  Similarly, this
prediction of the ADAF model may explain the peculiar spectral behavior of
GRO~J1655--40 during these observations when the apparent power-law component is steep
and poorly-determined.

\subsection{Disk and Power-law Component Fluxes and Source Luminosity}

Figures~5a--c are plots of the total unabsorbed flux (Fig.~5a), the unabsorbed
bolometric flux from the disk blackbody component (Fig.~5b) and the unabsorbed flux
from the power-law component (2--100~keV) (Fig.~5c).  Figure~5d shows the ratio of the
disk blackbody to the total flux.  During the flares, the power-law component
dominates the spectrum and we see that the increase in the power-law flux (Fig.~5c) is
accompanied by a decrease in the flux from the disk (Fig.~ 5b).  When the source moves
out of the solar exclusion zone, the power-law flux is less than  $5\times10^{-9}$ erg
s$^{-1}$ cm$^{-2}$ and the flux from the disk component dominates the spectrum.  This
behavior is even more apparent in Figure~5d, where the disk flux is plotted as a
fraction of the total unabsorbed flux from 2--100~keV.  During the hard flares, the
power-law dominates the unabsorbed flux, contributing $\sim$~70\% of the total flux
(Fig.~5d).  During the calm period (around day 160) between the two flaring episodes,
the power-law contributes less than 20\% of the flux (Fig.~5d).  After passage through
the SEZ, the disk emission dominates, contributing around 90\% of the total flux
(Fig.~5d).  After day 465, the power-law again dominates the unabsorbed flux,
contributing $\sim$~70\% of the total (Fig.~5d), indicating that the system has 
undergone a soft-to-hard transition.  

Figures~6a \& b show $r_{eff}$ and $\Gamma$ vs. the luminosity in Eddington units,
where $L_{Edd}=8.78\times10^{38}$ erg s$^{-1}$ for $m=7.02$.  Figure~6a indicates that
$r_{eff}$ decreases dramatically when the luminosity exceeds $l\sim0.17$ and Figure~6b
shows that the photon index is not strongly correlated with luminosity.  Both of these
results are unexpected for canonical outburst states, for which the parameters vary
smoothly (e.g. Ebisawa et al. 1994).  Also note from Figure~6 that the luminosity of
the source in each outburst state is several times lower than the canonical values
discussed in \S~1.  This data indicates that it is problematic to use absolute
luminosity as an indicator of spectral states.

\section{Correlations in the Spectral Parameters; Comparisons to Nova Mus 1991 and
GRS~1915$+$105}

Figures~3a \& b show that during the flares, $T_{eff}$ increases from $\sim$~0.76 to
1.13~keV, while $r_{eff}$ decreases by a factor of about three from $\sim$~5.5 to
1.6.  Figure~7a shows the correlation between $T_{eff}$ and $r_{eff}$, from which it
is apparent that $T_{eff}$ increases gradually as $r_{eff}$ decreases.  Observations
during the very high state are plotted as filled circles and observations during the
high/soft state are plotted as open squares.  Figure~7b shows a surprising correlation
between the power-law flux and $r_{eff}$.  From Figure~7b it is apparent that the
power-law flux is strongly correlated with $r_{eff}$ and increases steeply for
$r_{eff}\lesssim5.5$.  

Figures~8a \& b show $T_{eff}$ and the 2--100~keV power-law flux vs. $r_{eff}$ for
Nova Muscae 1991 (calculated from data in Ebisawa et al. 1994).  Comparing Figures~7b
\& 8b, it is apparent that both Nova Muscae 1991 and GRO~J1655--40 exhibit the same
behavior:  the power-law flux increases as $r_{eff}$ decreases.  

However, the apparent decrease of the inner disk radius observed during the power-law
flares may be caused by the failure of the multicolor disk model instead of physical
changes of the inner disk radius.  During the hard flares, the disk contributes as
little as 30\% of the observed flux.  Since the power-law component in BHXN spectra
likely originates from Compton upscattering of soft disk photons, an increase in
power-law emission naturally implies an increase in electron scattering.  This in turn
increases the distortion (spectral hardening) of the inner disk spectrum through
Comptonization of the emergent spectrum.  Therefore, we might expect the distortion of
the inner disk spectrum to be correlated with the power-law emission, as in Figures~7b
\& 8b.  

Shimura \& Takahara (1995) have shown that for $\alpha_v>0.1$ the inner disk gradually
becomes optically thin and the assumption of a constant spectral hardening factor
breaks down.  The spectral hardening factor ($f$) can become large (as high as
$f\sim10$ near the inner disk radius in an example given by Shimura \& Takahara) and
the color temperature can assume a steep radial profile.  In such a case, fitting the
multicolor disk model to the resulting spectrum and assuming a constant spectral
hardening factor yields an inner disk radius which is smaller than the physical value.
{\it The apparent decrease of the inner disk radius observed during periods of
increased power-law emission may be caused by a steepening of the radial
temperature profile of the accretion disk coupled with increased spectral hardening;
thus, this apparent decrease does not necessarily represent a physical change of the
inner disk radius}.  

This interpretation implies that the physical mechanism responsible for producing the
power-law emission is linked to the inner disk region and that the relativistic
electrons are produced very near the inner accretion disk.  Moreover, it is reasonable
that the hard X-ray spectrum will also be produced in the inner disk region since
there is a copious supply of soft seed photons there.  

GRS~1915+105 and GRO~J1655--40 are among the few Galactic sources known to produce
superluminal radio jets (Mirabel \& Rodriguez 1994, 1998; Tingay et al. 1995;
Hjellming \& Rupen 1995).  Although no dynamical mass estimate for the compact primary
in GRS~1915$+$105 exists, both sources are suspected of containing rapidly rotating
black holes (Zhang et al. 1997a; Cui et al. 1998).  Furthermore, GRS~1915$+$105 and
GRO~J1655--40 have exhibited large changes in the observed temperatures and radii of
their inner accretion disks.  Belloni et al. (1997a,b) explain the behavior of
GRS~1915$+$105 by the removal and replenishment of the matter forming the inner part
of an optically thick disk, probably caused by a Lightman-Eardley type thermal-viscous
instability (Lightman \& Eardley 1974) analogous to the limit cycle instability in
dwarf novae.  

The thermal instability model of Belloni et al. (1997a,b) describes the behavior of
GRS~1915$+$105 well, but it cannot account for the variability time scale of
GRO~J1655--40.  The flares in GRS~1915$+$105 occur on a time scale of $\sim100$
seconds, whereas the flares in GRO~J1655--40 occur on a time scale of a few days.  In
the Belloni et al. model, flares are expected to repeat on the time scale necessary to
refill the inner disk.  In the case of GRO~J1655--40, Belloni et al.'s model would
predict a flaring time scale of $\sim0.1$~seconds, which is some six orders of
magnitude shorter than the observed $\sim$~1-day flaring time scale.  Therefore, the
thermal instability model of Belloni et al. cannot be responsible for the flaring
behavior observed in GRO~J1655--40.  Instead, the time scale of the flaring behavior
in GRO~J1655--40 is comparable to the viscous time scale of the {\it outer} disk,
where the viscous time scale is $\sim1$~day for $r\sim300$.  This flaring time scale
is close to the 6-day X-ray delay (relative to the optical) that occurred during the
1996 April outburst of GRO~J1655--40, which Hameury et al. (1997) identified with the
viscous time scale needed to rebuild the inner disk.  Changes in the behavior of an
accreting system on this time scale can be associated with changes in the mass
accretion rate.

\section{Spin of the Black Hole}

Zhang et al. (1997a) outline general relativistic corrections to the inner disk radius
derived from the multicolor disk model for the purpose of determining the true inner
disk radius.  The angular momentum ($a_*$) of the black hole can be estimated by
equating the inner disk radius with the last stable orbit.  In this analysis, the
effective temperature of the disk peaks at a radius 
\begin{equation} 
r_{eff} = r_{last}/\eta, 
\end{equation} 
where $r_{last}$ is the last stable orbit and $\eta$ is a function of the black hole
angular momentum.  There is also a general relativistic correction to the spectral
hardening factor and an additional change of the observed flux due to viewing angle
(Zhang et al. 1997a).  However, the inclination angle of GRO~J1655--40 makes the
general relativistic spectral hardening and viewing angle corrections less than 10\%
(determined from data in Zhang et al. (1997a)) compared to the $\pm$24\% uncertainty
in the inner disk radius due to the conventional spectral hardening factor, $f = 1.7
\pm 0.2$ (see Eq.  (4)).  The important claim of Zhang et al.'s analysis is that {\it
$r_{eff}$ is the radius of the peak emission region and not the inner disk radius}.  

The last stable orbit around a maximally rotating prograde black hole ($a_*=+1$) can
extend almost to the event horizon, $GM/c^2=r_g$, whereas in the nonrotating
Schwarzschild case, the last stable orbit is at $6~r_g$.  The solutions for $r_{eff}$,
$r_{last}$, and $\eta$, for several values of $a_*$ calculated using the method of
Zhang et al. (1997a) are listed in Table~3.  It is important to note that this
analysis assumes the disk extends all the way down to the last stable orbit, and
therefore would not be valid throughout most of the observations presented here if the
inner disk radius were actually varying.  However, as emphasized in \S~4, the apparent
decrease of the inner disk radius observed during the power-law flares may be due
to the failure of the multicolor disk model and not to physical changes of the inner
disk radius.  

During the most stable period (from days 307--356), we find the average
effective inner disk radius is $r_{eff}=6.63\pm0.04$.  It is likely that during this
stable period, the disk assumes the steady state $r^{-3/4}$ temperature profile of the
standard multicolor disk model, which implies that $r_{eff}=6.6$ is a good
approximation of the actual effective inner disk radius.  Applying Zhang et al.'s
(1997a) method for $r_{eff}=6.6$ implies $r_{last}=4.2$ and $a_*=0.5$ (Table~3), which
is considerably smaller than the results of Zhang et al. who find $a_*=0.93$.  

The dominant source of uncertainty in this calculation (other than systematic
uncertainty in the physical interpretation of the disk model) is due to the spectral
hardening factor ($f=1.7\pm0.2$), which yields an uncertainty of 24\% in the value of
$r_{eff}$ (see Eq. (4)).  Taking this uncertainty into account, the minimum effective
inner disk radius $r_{eff}(min)=5.0$ corresponds to the upper limit $a_*<0.7$.  Since
$r_{eff}$ scales linearly with the assumed distance (see Eq. (1)), applying the firm
distance upper limit of 5~kpc from Tingay et al. (1995) and the 24\% uncertainty from
the spectral hardening factor, we obtain $r_{eff}(max)<12.9$, which is consistent with
a nonrotating (Schwarzschild) black hole (see Table~3).  

Using the method adopted here for measuring the inner disk radius, Zhang et al. 
(1997a) analyzed a single ASCA observation of GRO~J1655--40 from 1995 August, during a
previous outburst, and determined that the source contains a rapidly rotating black
hole with $a_*=0.93$, in contrast to our limit of $a_*<0.7$.  However, as demonstrated
in \S~4, the apparent inner disk radius can vary significantly during outburst.  This
result emphasizes the importance of obtaining good spectral coverage of the entire
outburst cycle.

\section{QPOs}

It has been proposed that high frequency QPOs in the X-ray lightcurves of BHXN are due
to the effects of relativistic frame dragging on an accretion disk orbiting a rotating
black hole (Cui et al.  1998).  According to this model, the 300~Hz QPO present during
the hardest spectral state in GRO~J1655--40 (Remillard et al. 1997), implies a
specific angular momentum for the black hole of $a_*=0.95$ for $M=7~M_{\odot}$ (Cui et
al. 1998).  The same model applied to the 67~Hz QPO observed in certain spectral
states of GRS~1915$+$105 (Morgan et al.  1997) also implies a specific angular
momentum for the black hole of $a_*=0.95$ for $M=30~M_{\odot}$ (Cui et al. 1998).  

In the previous section, we demonstrated that the most likely value of the inner disk
radius in GRO~J1655--40 implies $a_*<0.7$ and is consistent with a nonrotating
(Schwarzschild) black hole.  This result implies that the 300~Hz QPO in GRO~J1655--40
probably cannot be due to relativistic frame dragging, which requires a rapidly
rotating ($a_*\sim0.95$) black hole.  The low black hole angular momentum in
GRO~J1655--40 is also inconsistent with `diskoseismic' models in which the 300~Hz QPO
is attributed to g-mode or c-mode oscillations trapped in the inner accretion disk
(Nowak \& Wagoner 1992,1993; Perez et al. 1997), because these models also require a
rapidly rotating black hole with $a_*\sim0.95$ and $a_*\sim0.85$, respectively
(Wagoner 1998).  

Finally, we investigate the relationship between the derived spectral parameters and
the intermediate, variable frequency QPOs (14--28~Hz) in GRO~J1655--40.  Figures~9a \&
b illustrate a surprising correlation between QPO frequency, $r_{eff}$, and the
2--100~keV power-law flux for the variable frequency (14--28~Hz) QPOs.  Figure~9a
shows that the QPO frequency decreases as $r_{eff}$ decreases, which is contrary to
the behavior expected if the QPO were tied to the Keplerian frequency at the shrinking
inner edge of the disk.

\section{Summary \& Conclusion}

We have analyzed RXTE data obtained for GRO~J1655--40 covering a complete, 16-month
outburst cycle of the source.  These data comprise a nearly continuous observation of
the source with the ASM, and 52 pointed observations with the PCA and HEXTE
instruments.  Satisfactory fits to nearly all the PCA/HEXTE data were obtained with a
multicolor blackbody disk plus power-law model (with allowance for some minor
modifications).  The source exhibits two principal outburst states which we associate
with the very high and high/soft states.  During the very high state, the spectrum was
often dominated by a power-law component with $\Gamma =$~2.3--2.7.  The source
exhibited intense flares which were correlated with changes (hours to days) in both
the fitted temperature and radius of the inner accretion disk.  During the high/soft
state, the spectrum was dominated by the soft thermal emission from the accretion disk
with approximately constant inner disk radius and temperature.  The power-law
component was relatively weak with $\Gamma \sim$~2--3.  During the last few
observations, the source underwent a transition to the low/hard state.  

We found that a tight relationship exists between the observed inner radius of the
disk and the flux in the power-law component (Fig.~7b).  During intense hard flares,
the effective inner disk radius is observed to decrease by as much as a factor of
three (from 5.5 to 1.6~$r_g$) on a time scale of days.  The apparent decrease of the
inner disk radius observed in GRO~J1655--40 during periods of increased power-law
emission may be explained by the failure of the multicolor disk model due to
steepening of the radial temperature profile of the accretion disk coupled with
increased spectral hardening and not physical changes of the inner disk radius.  This
interpretation is consistent with the correlation between $r_{eff}$ and the power-law
flux (Fig.~7b) and implies that the physical mechanism responsible for producing the
power-law emission is linked to the inner disk region.  The same correlation between
the inner disk radius and power-law flux is also evident in the spectral results for
Nova Muscae 1991 presented by Ebisawa et al. (1994) and illustrated here (Fig.~8b).  

Assuming that our spectral model is valid during periods of weak power-law emission,
our most likely value for the inner disk radius implies $a_*<0.7$.  Such a low value
for the black hole angular momentum in GRO~J1655--40 is inconsistent with the
relativistic frame dragging and the `diskoseismic' models for the 300~Hz QPO, because
these models require a rapidly rotating black hole with $a_*\sim0.95$ and $a_*>0.85$,
respectively.

\acknowledgements

Partial support for J.M. and G.S. was provided by the Smithsonian Institution
Scholarly Studies Program.  R.R. was supported, in part, by NASA grant NAG5-3680. 
C.B. acknowleges support from NSF grant AST 9730774.  G.S. and J.M. wish to thank
D. Psaltis, A. Esin, S. Kenyon, R. Narayan, and J. Grindlay for helpful discussions, 
L. Titarchuk for his constructive correspondence, and an anonymous referee for
constructive criticism.  This research has made use of
data obtained through the High Energy Astrophysics Science Archive Research Center
Online Service, provided by the NASA/Goddard Space Flight Center.  

As we were completing the revision of this manuscript, Shahbaz et al. (MNRAS, in
press) presented additional radial velocities of GRO~J1655--40 obtained during
quiescence.  They derive an optical mass function of $f(M)=2.73 \pm 0.09~M_{\odot}$
(compared to $3.24 \pm 0.09~M_{\odot}$ reported by Orosz \& Bailyn (1997)).  The
implied black hole mass would then be $6.0 \pm 0.2~M_{\odot}$ (using the mass ratio
and inclination given by Orosz \& Bailyn (1997)), rather than $7.0 \pm 0.2~M_{\odot}$.
The main conclusions of this work (the correlation between fitted inner disk radius
and power-law flux, etc) are independent of the assumed black hole mass.  However,
since the effective radius of the inner disk scales inversely with the black hole
mass, a smaller black hole mass would imply a larger inner disk radius and hence a
lower angular momentum for the black hole (i.e. $a_*<0.6$ for $M=6~M_{\odot}$).

\newpage

\figcaption[fig1.eps]{(a) The 2--12~keV ASM lightcurve (counts s$^{-1}$) and (b) the
ratio of the ASM count rates (5--12~keV)/(3--5~keV) for GRO~J1655--40.  The labels in
the top panel and the dashed vertical lines indicate different outburst states (see
text).  The small, solid vertical lines in the top panel indicate the times of pointed
RXTE observations.  The individual ASM dwells are plotted in (a) and the ratios of the
one-day averages are plotted in (b).  \label{f1}}

\figcaption[fig2.eps]{Representative spectra from each of the observed spectral states
of GRO~J1655--40, along with the best fit model for (a, b) the very high state, (c, d)
the high/soft state, and (e) the low/hard state.  The individual components of the
model are also shown.  The 6-digit number in the upper left hand corner of each panel
identifies the date of the observation (see Table~1).  See the text for details on the
spectral models and fitting.  Although error bars are plotted for all the data, they
are only large enough to be visible at the highest energies.  \label{f2}}

\figcaption[fig3.eps]{Spectral parameters for PCA/HEXTE observations of GRO~J1655--40.
The spectra were fit to a model of interstellar absorption plus multicolor blackbody
disk plus power-law.  The hydrogen column density was fixed at $0.89 \times 10^{22}$
cm$^{-2}$ (Zhang et al. 1997b).  The quantities plotted here are (a) the effective
temperature of the accretion disk ($T_{eff}$) in keV, (b) the effective inner disk
radius $r_{eff}$ in units of $r_g = G M/c^2$ for $M = 7.02 M_{\odot}$, $\theta = 69
\fdg 5$ (Orosz \& Bailyn 1997), and $D = 3.2$ kpc (Hjellming \& Rupen 1995) (see Eq. 
(1)), (c) the power-law photon index $\Gamma$, and (d) the power-law normalization $K$
in units of photons s$^{-1}$ cm$^{-2}$ keV$^{-1}$ at 1~keV.  The presence of the
300~Hz QPO is indicated by an open triangle and the presence of the variable frequency
QPOs (14--28~Hz) is indicated by an open square at the top of (b) (Remillard et al.
1998).  Here and in several subsequent figures for which error bars are not visible,
it is because they are comparable to or smaller than the plotting symbol.  \label{f3}}

\figcaption[fig4.eps]{Effect of the Fe absorption edge on the ratio data/model for a
representative high/soft state spectrum (970320).  Figure~(a) shows the residuals for
the interstellar absorption plus multicolor blackbody disk plus power-law only. 
Figures~(b) \& (c) show the effects of a smeared Fe absorption edge applied to the
power-law component (Ebisawa et al. 1994) at edge energies of 7.1 and 8.0~keV,
respectively.  The width of the absorption feature was kept fixed at 7~keV and the Fe
optical depth was allowed to float.  \label{f4}}

\figcaption[fig5.eps]{Plot of (a) the total unabsorbed flux of GRO~J1655--40, (b) the
bolometric flux from the disk blackbody calculated using the formula: $Flux = 2 \sigma
\left( (R_{col}^2/D^2) \cos{\theta} \right) T_{col}^4$, and (c) the 2--100~keV flux
from the power-law component.  The units are 10$^{-8}$ ergs s$^{-1}$ cm$^{-2}$.  The
ratio of the disk blackbody to the total unabsorbed flux is shown in (d).  The total
and power-law fluxes for the ten poorly-fit high/soft state spectra have been omitted
because of uncertainty in the nature of the quasi-power-law component.  Reducing the
lower energy bound from 2~keV to 1~keV increases the power-law fluxes by $\sim$~50\%. 
\label{f5}}

\figcaption[fig6.eps]{Plot of (a) $r_{eff}$ vs. $l$ and (b) $\Gamma$ vs. $l$, where
$l$ is the bolometric disk luminosity plus the 2--100~keV power-law luminosity in
Eddington units ($L_{Edd} = 8.8 \times 10^{38}$~erg~s$^{-1}$ for $M = 7.02
M_{\odot}$).  Data during the very high state are plotted as filled circles and data
during the high/soft state are plotted as open squares.  The last three data points on
days 476, 480, and 487 have been omitted from (a) because the multicolor disk model no
longer yields reasonable physical values for the inner disk radius.  These three data
points are plotted as open triangles in (b).  The data points for the ten poorly-fit
high/soft state spectra have been omitted from both (a) \& (b) because of uncertainty
in the nature of the quasi-power-law component.  \label{f6}}

\figcaption[fig7.eps]{Plot of (a) $T_{eff}$ vs. $r_{eff}$ and (b) 2--100~keV power-law
flux vs. $r_{eff}$.  Observations during the very high state are plotted as filled
circles and observations during the high/soft state are plotted as open squares.  The
last three observations on days 476, 480, and 487 have been omitted because the
multicolor disk model no longer yields reasonable physical values for the inner disk
radius.  The data points for the ten poorly-fit high/soft state spectra have been
omitted because of uncertainty in the nature of the quasi-power-law component. 
\label{f7}}

\figcaption[fig8.eps]{Plot of (a) $T_{eff}$ vs. $r_{eff}$ and (b) 2--100~keV power-law
flux vs. $r_{eff}$ for Nova Muscae 1991.  This figure should be compared to Figure~7
for GRO~J1655--40.  $r_{eff}$ is in units of $r_g = G M/c^2$ where $M = 6 M_{\odot}$,
$\theta = 60^{\circ}$ (Orosz et al. 1996), and $D = 2.5$~kpc (Ebisawa et al. 1994). 
Following Ebisawa et al. (1994) and for the purposes of this illustration we adopt $D
= 2.5$~kpc.  However, the actual distance is probably closer to twice this value (see
\S~4.1 in Esin et al. 1997), which would double the value of $r_{eff}$.  \label{f8}}

\figcaption[fig9.eps]{Correlation between (a) QPO frequency and $r_{eff}$ and (b) QPO
frequency and the 2--100~keV power-law flux for 14--28~Hz QPOs.  The 28.3~Hz QPO
plotted with a triangle appears only in the sum of the high/soft state observations
(days 255--465) (Remillard et al. 1998), and is plotted here at the average $r_{eff}$
and power-law flux for the high/soft state.  \label{f9}}

\newpage
\singlespace

\begin{deluxetable}{ccccccccc}
\scriptsize
\tablenum{1}
\tablecaption{Spectral Parameters for GRO J1655$-$40\tablenotemark{a}}
\tablehead{
 \colhead{Date} & \colhead{Day\tablenotemark{b}} & \colhead{$T_{eff}$ (keV)} & 
 \colhead{$r_{eff}$ ($r_g$)\tablenotemark{c}} & \colhead{Photon} &
 \colhead{Power-law Norm.} & \colhead{$\tau_{Fe}$\tablenotemark{d}} & \colhead{$\chi^2_{\nu}$} & \colhead{Energy} \cr 
 \colhead{(yymmdd} & \colhead{(MJD} & \colhead{ } & 
 \colhead{ } & \colhead{Index} & 
 \colhead{(phot s$^{-1}$ cm$^{-2}$} & \colhead{ } & \colhead{(dof)} & \colhead{Range\tablenotemark{e}} \cr 
 \colhead{UT)} & \colhead{50198$+$)} & \colhead{ } & 
 \colhead{ } & \colhead{ } & 
 \colhead{keV$^{-1}$ at 1 keV)} & \colhead{ } & \colhead{ } & \colhead{(keV)}}
\startdata
960509&   14.8& $0.73 \pm 0.01$\tablenotemark{f}&   $6.42 \pm 0.04$\tablenotemark{f}&    $5.77_{-0.19}^{+0.21}$\tablenotemark{g}&   $4620_{-1900}^{+3640}$\tablenotemark{g}&    $-$&   1.65(39)\tablenotemark{f}&   $-$\nl
960510&   15.5& $0.72 \pm 0.01$\tablenotemark{f}&   $6.57 \pm 0.04$\tablenotemark{f}&    $6.47_{-0.20}^{+0.33}$\tablenotemark{g}&   $31400_{-13100}^{+47100}$\tablenotemark{g}&   $-$&   1.13(39)\tablenotemark{f}&   $-$\nl
960511&   16.5& $0.73 \pm 0.01$\tablenotemark{f}&   $6.48 \pm 0.04$\tablenotemark{f}&    $5.69_{-0.26}^{+0.30}$\tablenotemark{g}&   $3930_{-2030}^{+5210}$\tablenotemark{g}&    $-$&   1.45(39)\tablenotemark{f}&   $-$\nl
960512&   17.4& $0.73 \pm 0.01$\tablenotemark{f}&   $6.36 \pm 0.05$\tablenotemark{f}&    $5.75_{-0.26}^{+0.32}$\tablenotemark{g}&   $4900_{-2530}^{+7140}$\tablenotemark{g}&    $-$&   1.13(39)\tablenotemark{f}&   $-$\nl
960725&   91.4& $0.77 \pm 0.01$&   $5.37 \pm 0.04$&    $2.28 \pm 0.01$&   $5.90 \pm 0.18$&    $-$&   1.57(248)&   2.5-100\nl
960801\tablenotemark{h}&   98.4& $1.13 \pm 0.01$&   $1.64 \pm 0.06$&    $2.68 \pm 0.01$&   $62.7 \pm 1.3$&    $-$&   1.35(246)&   2.5-100\nl
960806&   103.7& $0.93 \pm 0.01$&   $2.70 \pm 0.05$&    $2.67 \pm 0.01$&   $44.1 \pm 0.9$&    $-$&   0.84(248)&   2.5-100\nl
960815&   112.6& $0.76 \pm 0.01$&   $5.35 \pm 0.04$&    $2.30 \pm 0.02$&   $5.90_{-0.35}^{+0.37}$&    $-$&   2.11(150)&   2.5-50\nl
960816&   113.4& $0.78 \pm 0.01$&   $4.49 \pm 0.07$&    $2.42 \pm 0.01$&   $18.9 \pm 0.6$&    $-$&   1.23(248)&   2.5-100\nl
960822&   119.5& $0.76 \pm 0.01$&   $4.96 \pm 0.05$&    $2.46 \pm 0.01$&   $16.4 \pm 0.5$&    $-$&   0.88(248)&   2.5-100\nl
960829\tablenotemark{h}&   126.4& $1.09 \pm 0.01$&   $1.98_{-0.09}^{+0.07}$&    $2.78 \pm 0.03$&   $65.5_{-2.2}^{+2.4}$&    $-$&   0.73(246)&   2.5-50\nl
960904&   132.3& $0.76 \pm 0.01$&   $5.34 \pm 0.05$&    $2.40 \pm 0.02$&   $9.59_{-0.41}^{+0.43}$&    $-$&   1.89(150)&   2.5-50\nl
960909&   138.0& $0.78 \pm 0.01$&   $5.07 \pm 0.05$&    $2.37 \pm 0.01$&   $9.49 \pm 0.22$&    $-$&   1.50(248)&   2.5-100\nl
960920&   148.2& $0.75 \pm 0.01$&   $5.44 \pm 0.04$&    $2.23 \pm 0.02$&   $4.31_{-0.22}^{+0.23}$&    $-$&   1.60(150)&   2.5-50\nl
960926&   154.3& $0.75 \pm 0.01$&   $5.72 \pm 0.04$&    $1.97 \pm 0.03$&   $0.93_{-0.06}^{+0.07}$&    $-$&   1.32(150)&   2.5-50\nl
961003&   161.6& $0.75 \pm 0.01$&   $5.65 \pm 0.04$&    $2.14 \pm 0.02$&   $2.16_{-0.12}^{+0.13}$&    $-$&   0.82(150)&   2.5-50\nl
961015&   173.5& $0.78 \pm 0.01$&   $5.39 \pm 0.04$&    $2.18 \pm 0.01$&   $3.83 \pm 0.13$&    $-$&   1.09(248)&   2.5-100\nl
961022&   180.1& $0.77 \pm 0.01$&   $5.55 \pm 0.05$&    $2.23 \pm 0.01$&   $4.86 \pm 0.15$&    $-$&   0.92(248)&   2.5-100\nl
961027&   185.6& $0.80 \pm 0.01$&   $4.56 \pm 0.06$&    $2.59 \pm 0.02$&   $26.0 \pm 1.5$&    $-$&   0.60(91)&   2.5-20\nl
961102A\tablenotemark{h}&  191.2& $0.84_{-0.02}^{+0.01}$&   $2.15_{-0.09}^{+0.10}$&    $2.57 \pm 0.01$&   $35.9 \pm 0.8$&    $-$&   1.26(246)&   2.5-100\nl
961102B\tablenotemark{h}&  191.3& $1.09 \pm 0.01$&   $1.69_{-0.09}^{+0.07}$&    $2.64 \pm 0.01$&   $53.4_{-1.5}^{+1.9}$&    $-$&   1.00(246)&   2.5-100\nl
970105&   255.4& $0.64 \pm 0.01$&   $5.72_{-0.06}^{+0.07}$&    $1.94 \pm 0.04$&   $0.92_{-0.12}^{+0.13}$&    $0.40 \pm 0.25$&   1.08(149)&   2.5-50\nl
970112&   262.1& $0.62 \pm 0.01$&   $6.27 \pm 0.07$&    $2.00 \pm 0.06$&   $0.61 \pm 0.13$&    $0.97 \pm 0.33$&   0.88(149)&   2.5-50\nl
970121&   271.0& $0.55 \pm 0.01$&   $6.50 \pm 0.05$&    $1.65 \pm 0.09$&   $0.03 \pm 0.01$&    $0.00_{-0.00}^{+2.97}$&   0.82(91)&   2.5-20\nl
970126&   276.9& $0.57 \pm 0.01$&   $6.65 \pm 0.10$&    $2.17 \pm 0.10$&   $0.48_{-0.13}^{+0.16}$&    $1.79_{-0.48}^{+0.42}$&   0.56(90)&   2.5-20\nl
970226&   307.9& $0.65 \pm 0.01$&   $6.62 \pm 0.05$&    $3.01 \pm 0.08$&   $4.50_{-0.92}^{+1.06}$&    $10.2 \pm 0.4$&   2.53(107)&   2.5-30\nl
970305&   314.8& $0.66 \pm 0.01$&   $6.70 \pm 0.06$&    $2.96 \pm 0.08$&   $4.89_{-1.00}^{+1.15}$&    $7.23 \pm 0.32$&   2.20(90)&   2.5-20\nl
970310&   319.7& $0.66 \pm 0.01$&   $6.61 \pm 0.05$&    $2.75 \pm 0.08$&   $3.46_{-0.76}^{+0.89}$&    $5.64_{-0.31}^{+0.29}$&   1.53(90)&   2.5-20\nl
970320&   329.9& $0.67 \pm 0.01$&   $6.63 \pm 0.04$&    $2.43 \pm 0.07$&   $1.99_{-0.42}^{+0.50}$&    $3.62_{-0.32}^{+0.30}$&   1.04(147)&   2.5-49\nl
970324&   333.8& $0.67 \pm 0.01$\tablenotemark{f}&   $6.61 \pm 0.03$\tablenotemark{f}&    $2.70_{-0.35}^{+0.40}$\tablenotemark{g}&   $0.61_{-0.61}^{+1.26}$\tablenotemark{g}&    $-$&   2.58(39)\tablenotemark{f}&   $-$\nl
970404&   344.7& $0.67 \pm 0.01$&   $6.64 \pm 0.04$&    $2.41 \pm 0.09$&   $1.88_{-0.45}^{+0.54}$&    $3.76_{-0.35}^{+0.32}$&   1.17(149)&   2.5-50\nl
970410&   350.5& $0.67 \pm 0.01$&   $6.58 \pm 0.04$&    $2.44 \pm 0.09$&   $2.38_{-0.60}^{+0.71}$&    $3.53_{-0.33}^{+0.30}$&   0.86(90)&   2.5-20\nl
970416&   356.8& $0.66 \pm 0.01$&   $6.65 \pm 0.05$&    $2.96 \pm 0.08$&   $4.19_{-0.91}^{+1.10}$&    $9.03_{-0.39}^{+0.38}$&   2.29(89)&   2.5-20\nl
970424&   364.8& $0.67 \pm 0.01$\tablenotemark{f}&   $6.58 \pm 0.03$\tablenotemark{f}&    $2.96 \pm 0.65$\tablenotemark{g}&   $1.14_{-1.14}^{+6.05}$\tablenotemark{g}&    $-$&   1.92(39)\tablenotemark{f}&   $-$\nl
970430&   370.6& $0.68 \pm 0.01$\tablenotemark{f}&   $6.50 \pm 0.03$\tablenotemark{f}&    $3.50 \pm 0.56$\tablenotemark{g}&   $4.76_{-4.76}^{+18.5}$\tablenotemark{g}&    $-$&   2.33(39)\tablenotemark{f}&   $-$\nl
970508&   378.5& $0.70 \pm 0.01$\tablenotemark{f}&   $6.40 \pm 0.03$\tablenotemark{f}&    $2.87_{-0.38}^{+0.42}$\tablenotemark{g}&   $1.07_{-1.07}^{+2.47}$\tablenotemark{g}&    $-$&   2.04(39)\tablenotemark{f}&   $-$\nl
970512&   382.7& $0.70 \pm 0.01$\tablenotemark{f}&   $6.40 \pm 0.03$\tablenotemark{f}&    $5.10_{-0.62}^{+0.55}$\tablenotemark{g}&   $387_{-318}^{+1400}$\tablenotemark{g}&    $-$&   1.90(39)\tablenotemark{f}&   $-$\nl
970520&   390.4& $0.71 \pm 0.01$&   $6.31 \pm 0.06$&    $2.89_{-0.12}^{+0.10}$&   $3.86_{-1.18}^{+1.37}$&    $4.82_{-0.57}^{+0.51}$&   1.43(90)&   2.5-20\nl
970528&   398.4& $0.71 \pm 0.01$&   $6.30 \pm 0.05$&    $2.52 \pm 0.08$&   $3.08_{-0.68}^{+0.80}$&    $2.92_{-0.35}^{+0.33}$&   0.91(147)&   2.5-49\nl
970605&   406.3& $0.69 \pm 0.01$&   $6.38 \pm 0.06$&    $2.86 \pm 0.09$&   $4.71_{-1.16}^{+1.30}$&    $4.73_{-0.20}^{+0.36}$&   1.10(108)&   2.5-30\nl
970609&   410.5& $0.68 \pm 0.01$&   $6.49 \pm 0.07$&    $2.89 \pm 0.11$&   $4.24_{-1.24}^{+1.47}$&    $6.59_{-0.51}^{+0.47}$&   1.10(90)&   2.5-20\nl
970619&   420.3& $0.67 \pm 0.01$&   $6.56 \pm 0.04$&    $2.24 \pm 0.11$&   $1.38_{-0.40}^{+0.47}$&    $3.39_{-0.40}^{+0.33}$&   0.80(90)&   2.5-20\nl
970626&   427.8& $0.65 \pm 0.01$\tablenotemark{f}&   $6.71 \pm 0.03$\tablenotemark{f}&    $3.35_{-0.73}^{+0.76}$\tablenotemark{g}&   $3.13_{-3.13}^{+23.3}$\tablenotemark{g}&    $-$&   0.95(39)\tablenotemark{f}&   $-$\nl
970704&   435.5& $0.61 \pm 0.01$&   $6.95 \pm 0.04$&    $2.64 \pm 0.09$&   $1.82_{-0.44}^{+0.52}$&   $6.10_{-0.30}^{+0.27}$&   1.51(90)&   2.5-20\nl
970708&   439.6& $0.62 \pm 0.01$&   $6.59 \pm 0.06$&    $2.10 \pm 0.10$&   $0.71_{-0.20}^{+0.23}$&    $2.10_{-0.45}^{+0.38}$&   0.79(90)&   2.5-20\nl
970714&   445.3& $0.60 \pm 0.01$&   $7.11 \pm 0.05$&    $2.53 \pm 0.11$&   $1.31_{-0.38}^{+0.47}$&    $5.13_{-0.39}^{+0.35}$&   0.87(90)&   2.5-20\nl
970724&   455.6& $0.57 \pm 0.01$&   $6.60 \pm 0.08$&    $2.16 \pm 0.05$&   $0.67_{-0.10}^{+0.12}$&    $1.94_{-0.24}^{+0.23}$&   1.09(149)&   2.5-50\nl
970729&   460.4& $0.51 \pm 0.01$&   $6.89 \pm 0.12$&    $2.22 \pm 0.05$&   $0.94_{-0.12}^{+0.13}$&    $2.32 \pm 0.25$&   0.88(90)&   2.5-20\nl
970803&   465.7& $0.46 \pm 0.01$&   $7.31 \pm 0.12$&    $2.52 \pm 0.06$&   $0.87_{-0.12}^{+0.13}$&    $2.19 \pm 0.26$&   1.01(90)&   2.5-20\nl
970814&   476.5& $0.31 \pm 0.02$&   $5.57_{-0.77}^{+0.94}$&    $2.00 \pm 0.01$&   $0.76 \pm 0.02$&     $0.85 \pm 0.07$&   0.87(149)&   2.5-50\nl
970818&   480.6& $0.33 \pm 0.04$&   $1.82_{-0.54}^{+0.94}$&    $1.71 \pm 0.02$&   $0.14 \pm 0.01$&     $0.34 \pm 0.10$&   0.78(119)&   2.5-50\nl
970825&   487.5& $0.26 \pm 0.03$&   $3.12_{-0.87}^{+1.45}$&    $1.89 \pm 0.11$&   $0.02 \pm 0.01$&     $1.20_{-0.52}^{+0.53}$&   0.51(74)&   2.5-17\nl
\enddata
\tablenotetext{a}{Used fixed $N_H = 0.89 \times 10^{22}$ cm$^{-2}$ (Zhang et al. 1997).}
\tablenotetext{b}{Midpoint of observation.}
\tablenotetext{c}{$r_g = G M/c^2$ for $M = 7.02 M_{\odot}$ and Eq. (1) with $\theta = 69 \fdg 5$ 
(Orosz \& Bailyn 1997) and $D = 3.2$ kpc (Hjellming \& Rupen 1995).}
\tablenotetext{d}{Fixed edge at 8.0 keV and width of 7 keV using `smedge' 
model in XSPEC.}
\tablenotetext{e}{PCA = $2.5 - 20$ keV \& HEXTE $>20$ keV.}
\tablenotetext{f}{Determined from 2.5-10 keV using only interstellar absorption 
and multicolor disk blackbody models.}
\tablenotetext{g}{Determined from 15-20 keV with fixed multicolor disk 
parameters determined from 2.5-10 keV.}
\tablenotetext{h}{Fitted including a Compton reflection component model (see \S 3.1 and
Table 2).}
\end{deluxetable}

\begin{deluxetable}{cccc}
\tablenum{2}
\tablecaption{Fit Parameters for Compton Reflection\tablenotemark{a}}
\tablehead{
 \colhead{Date} & \colhead{Day\tablenotemark{b}} & 
 \colhead{$\Omega/2\pi$\tablenotemark{c}} & \colhead{$\xi$\tablenotemark{d}} \cr 
 \colhead{(yymmdd} & \colhead{(MJD} & 
 \colhead{ } & \colhead{ } \cr 
 \colhead{UT)} & \colhead{50198$+$)} &
 \colhead{ } & \colhead{ }}
\startdata
960801&   98.4&   $0.88_{-0.08}^{+0.09}$&  $442_{-272}^{+499}$\nl
960829&   126.4&  $1.30_{-0.35}^{+0.27}$&  $84.5_{-79.2}^{+642}$\nl
961102A&  191.2&  $0.58 \pm 0.07$&         $1.01_{-0.97}^{+12.0}$\nl
961102B&  191.3&  $1.15_{-0.10}^{+0.11}$&  $0.013_{-0.013}^{+41.4}$\nl
\enddata
\tablenotetext{a}{The Compton reflection component was calculated using the `pexriv'
model in XSPEC version 10.  The Fe and elemental abundances were fixed at the solar
value.  The disk temperature was fixed at $10^5$ K (Done et al. 1992), the inclination
angle was fixed at $69 \fdg 5$ (Orosz \& Bailyn 1997).  No exponential cutoff of the
power-law was applied.} 
\tablenotetext{b}{Midpoint of observation.}
\tablenotetext{c}{Normalization of reflection.}
\tablenotetext{d}{The ionization parameter $\xi = L/n R^2$, where $L$ is the integrated
incident luminosity between 5 eV and 300 keV, $n$ is the density of the material, and $R$
is the distance of the material from the illuminating source (Done et al. 1992).}
\end{deluxetable}

\begin{deluxetable}{cccc}
\tablenum{3}
\tablecaption{Determining $r_{last}$ \& $a_*$ from $r_{eff}$}
\tablehead{
 \colhead{$r_{eff}$ ($r_g$)\tablenotemark{\dag}} & 
 \colhead{$r_{last}$ ($r_g$)\tablenotemark{\dag}} & 
 \colhead{$\eta$} & \colhead{$a_*$}}
\startdata
1.58& 1.24& 0.782& 0.998\nl 
1.66& 1.28& 0.771& 0.997\nl
1.75& 1.33& 0.760& 0.996\nl 
1.98& 1.45& 0.736& 0.99\nl
2.80& 1.94& 0.692& 0.95\nl 
3.08& 2.10& 0.683& 0.93\nl 
3.44& 2.32& 0.675& 0.90\nl 
4.42& 2.91& 0.658& 0.80\nl 
5.22& 3.39& 0.650& 0.70\nl 
5.94& 3.83& 0.645& 0.60\nl 
6.61& 4.23& 0.640& 0.50\nl 
7.24& 4.61& 0.637& 0.40\nl 
7.85& 4.98& 0.634& 0.30\nl 
8.43& 5.33& 0.632& 0.20\nl 
9.00& 5.67& 0.630& 0.10\nl
9.55& 6.00& 0.628& 0.00\nl 
\enddata
\tablenotetext{\dag}{$r_g = G M/c^2$}
\end{deluxetable}

\newpage
\begin{figure}
\figurenum{1}
\plotone{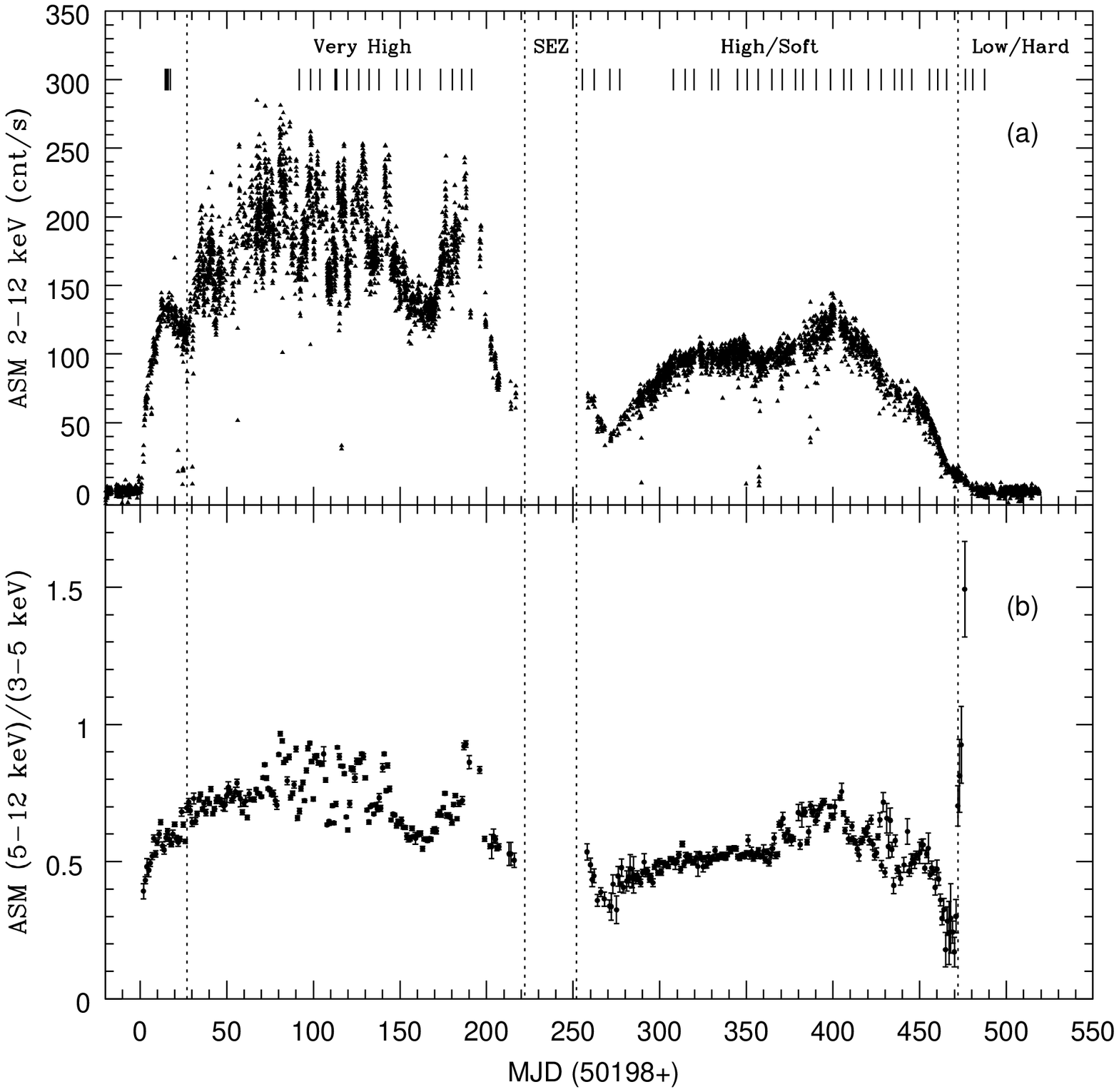}
\caption{ }
\end{figure}

\begin{figure}
\figurenum{2}
\plotone{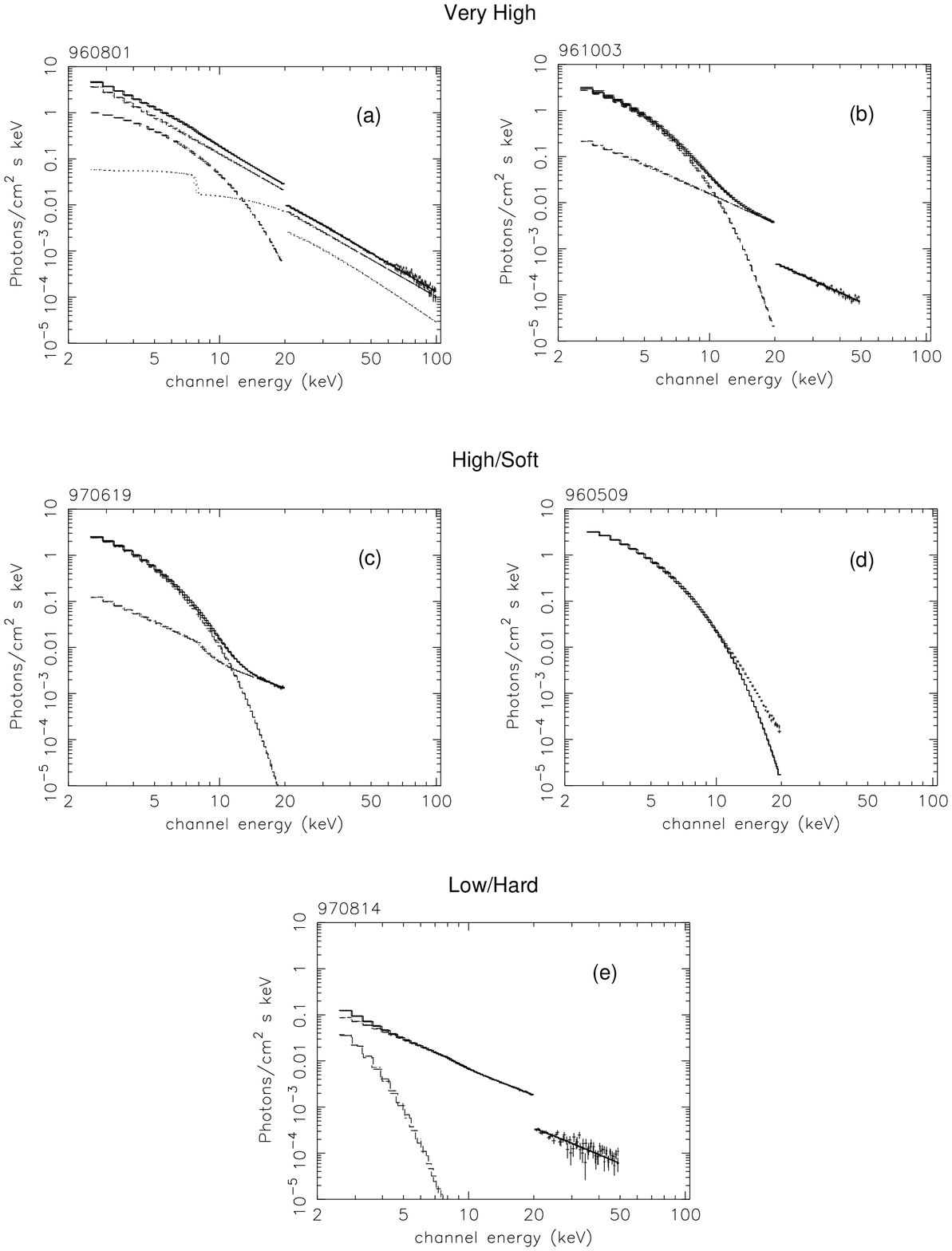}
\caption{ }
\end{figure}

\begin{figure}
\figurenum{3}
\plotone{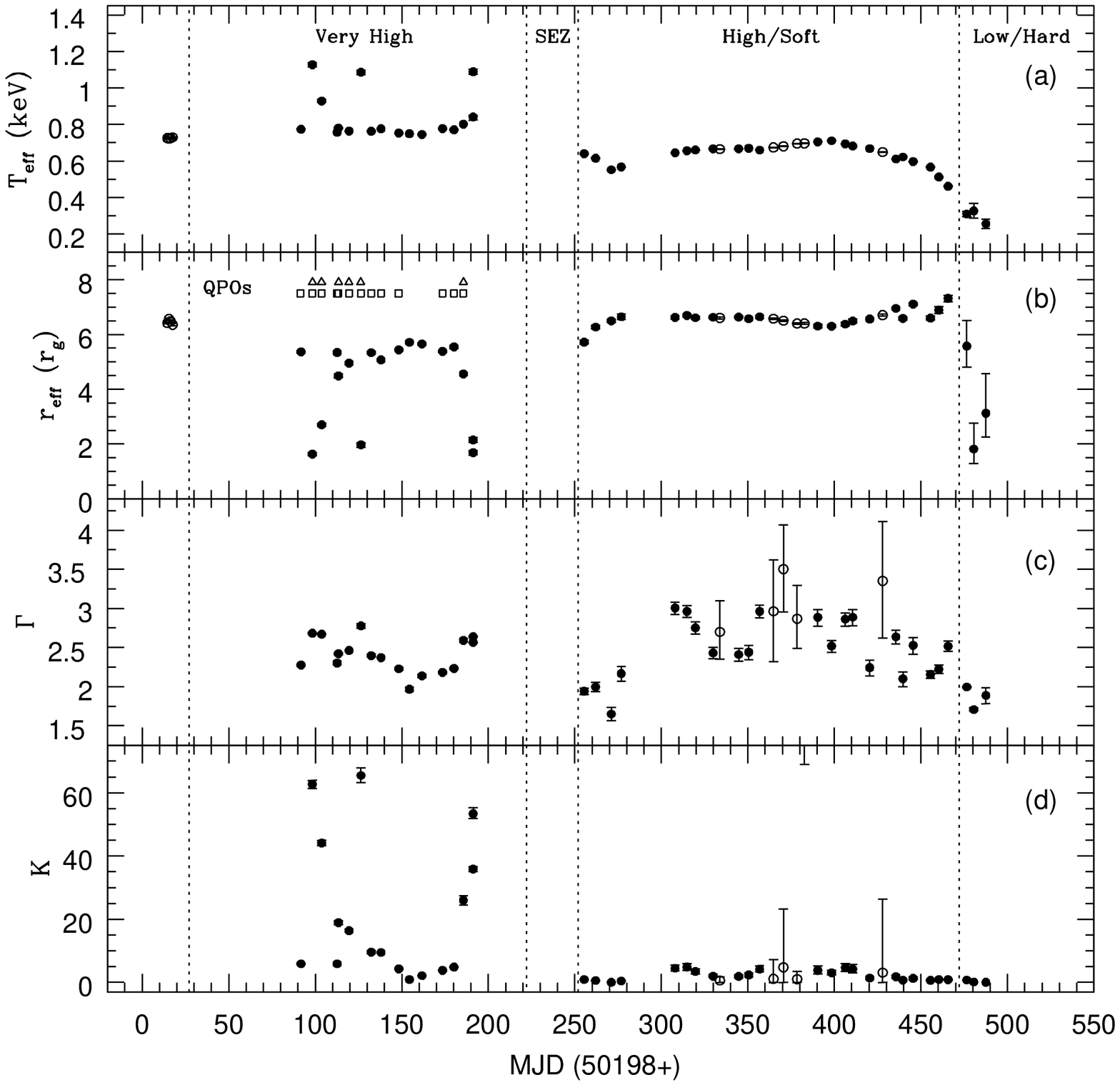}
\caption{ }
\end{figure}

\begin{figure}
\figurenum{4}
\plotone{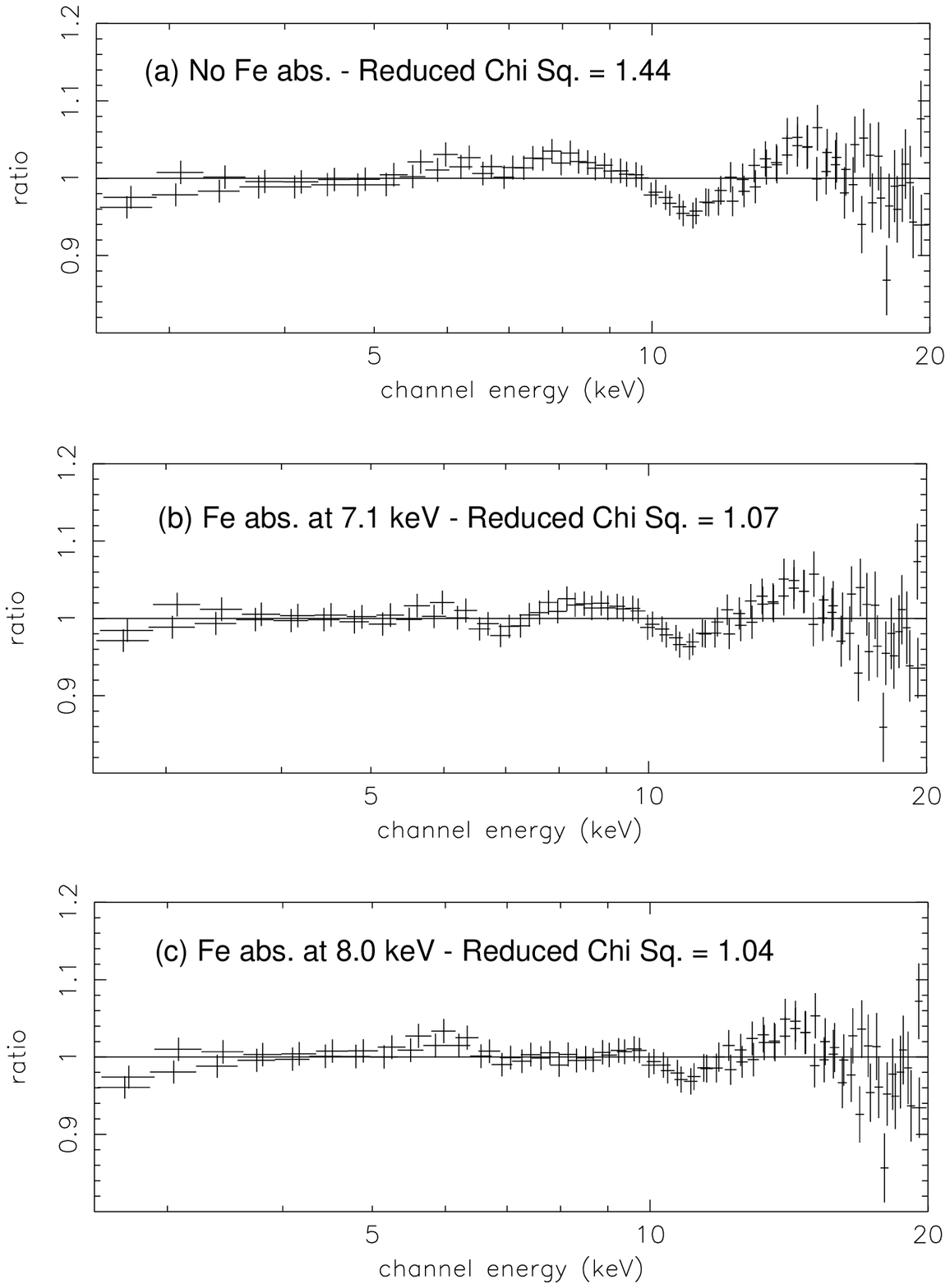}
\caption{ }
\end{figure}

\begin{figure}
\figurenum{5}
\plotone{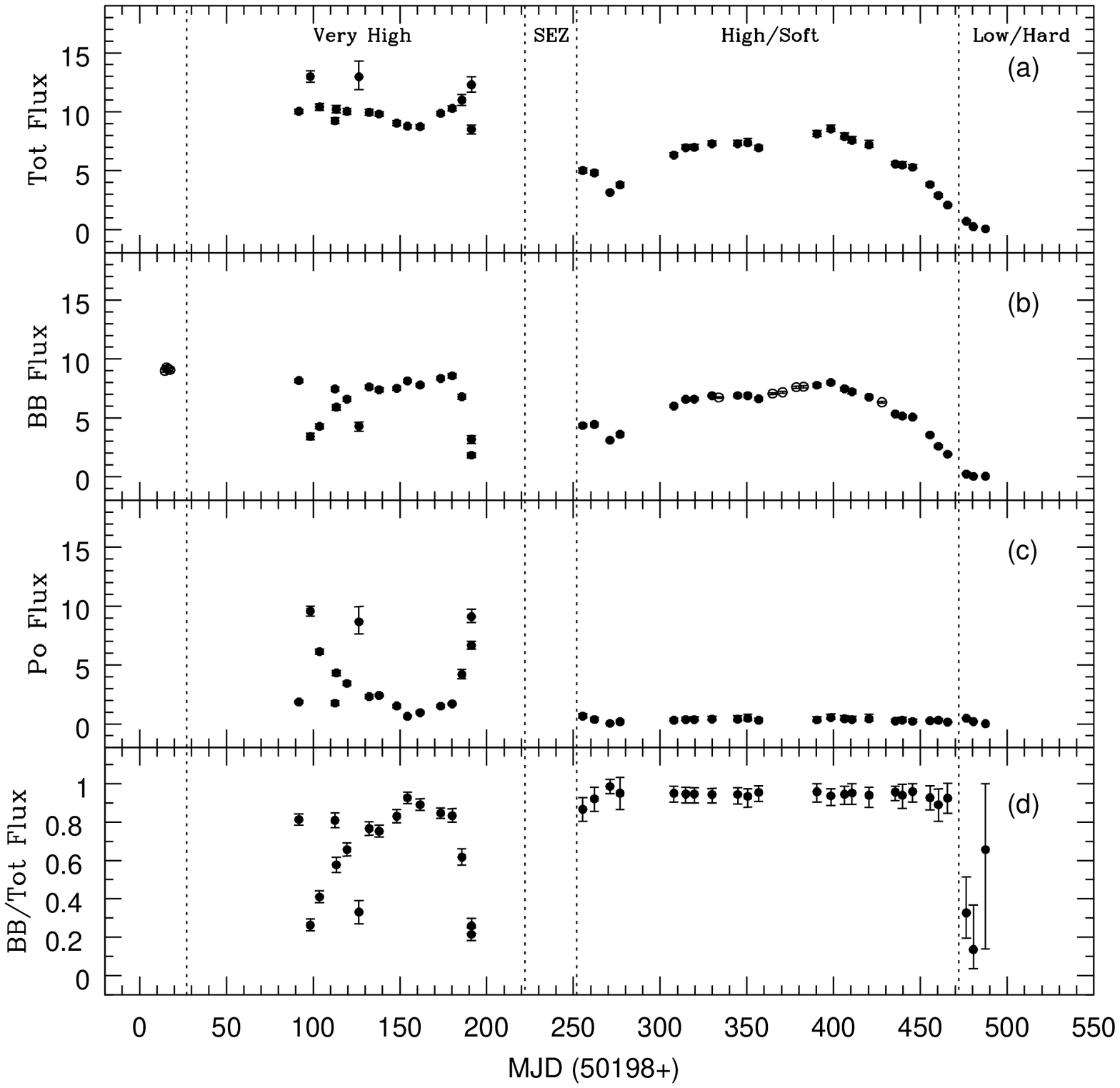}
\caption{ }
\end{figure}

\begin{figure}
\figurenum{6}
\plotone{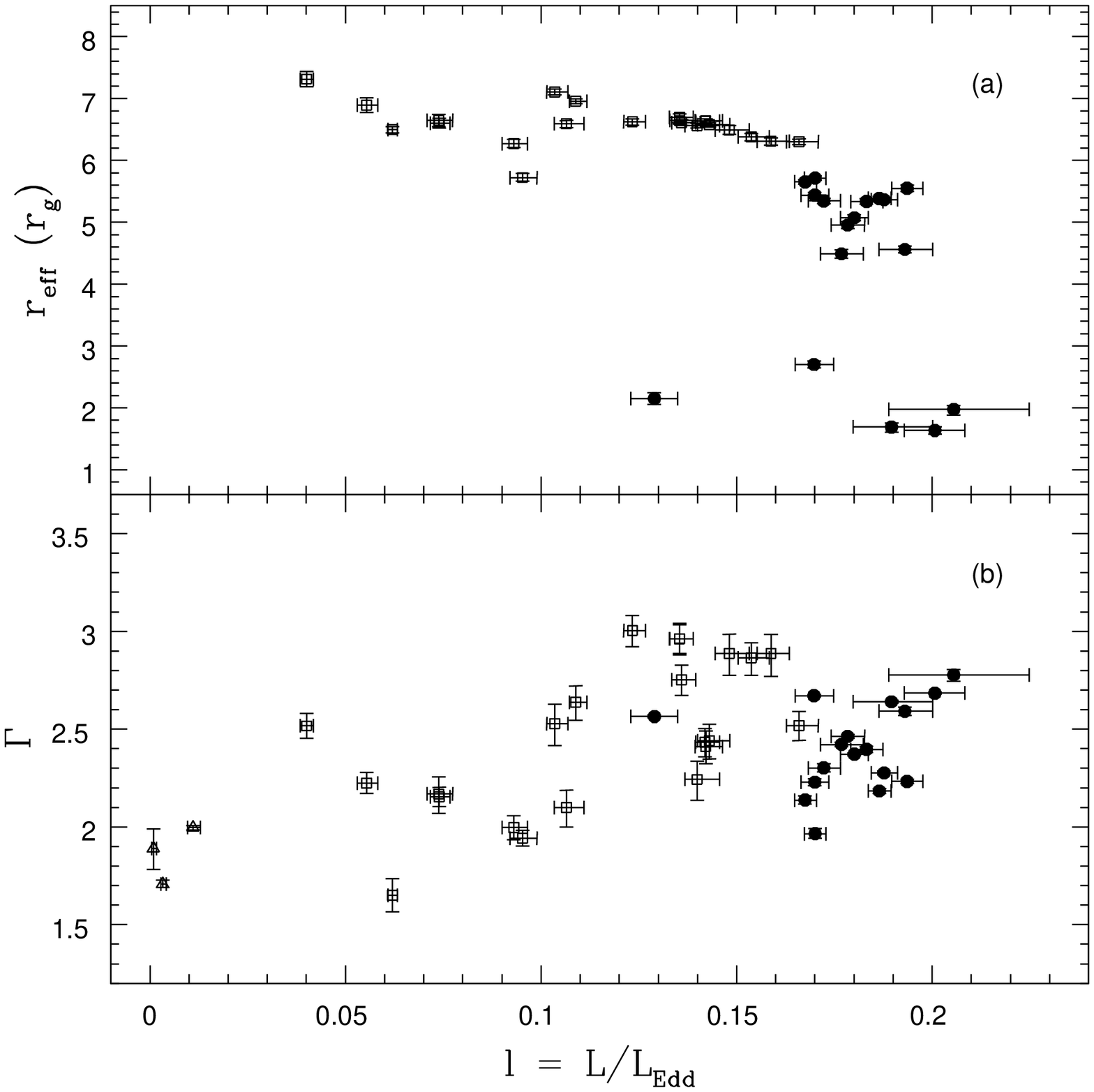}
\caption{ }
\end{figure}

\begin{figure}
\figurenum{7}
\plotone{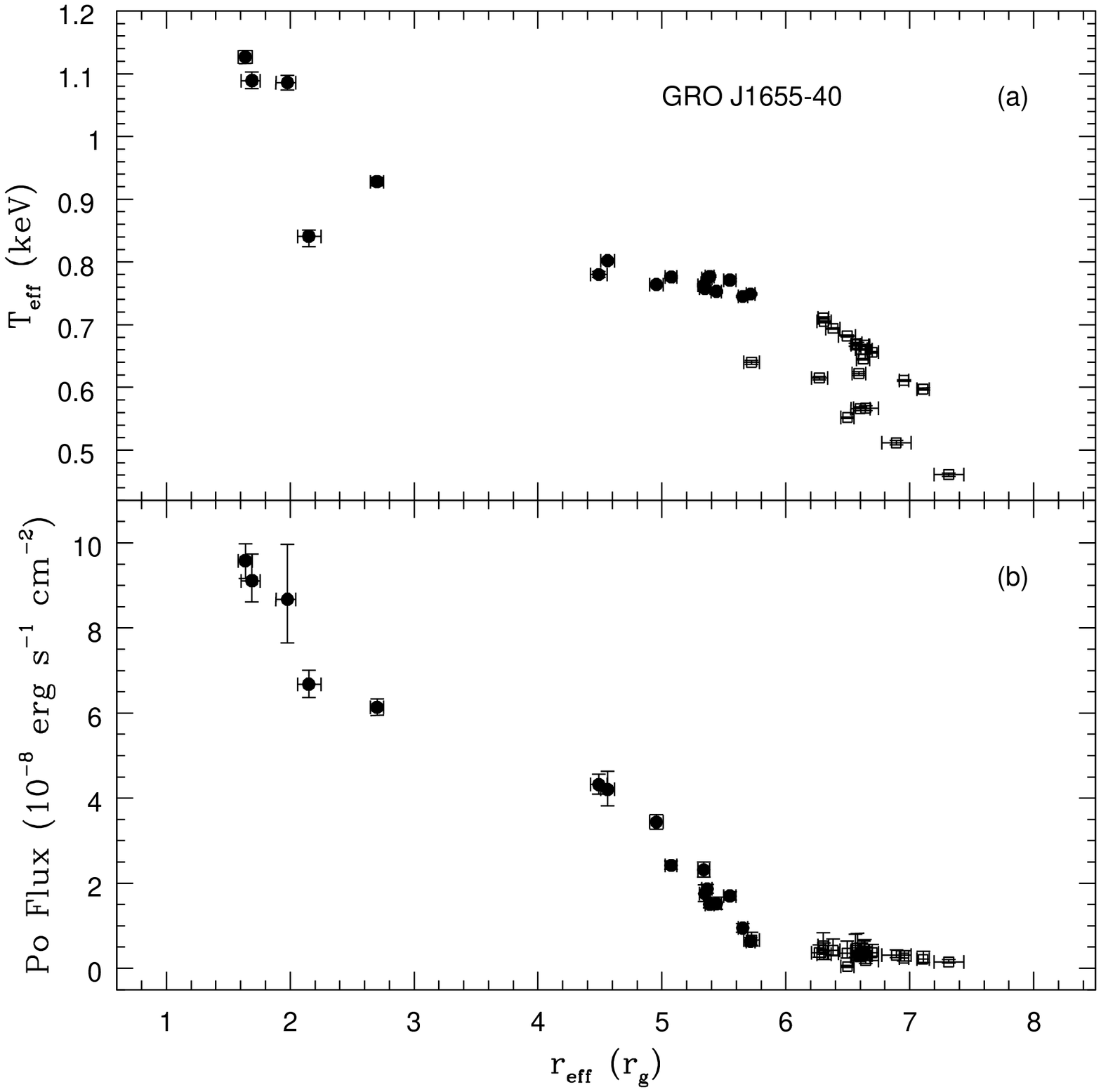}
\caption{ }
\end{figure}

\begin{figure}
\figurenum{8}
\plotone{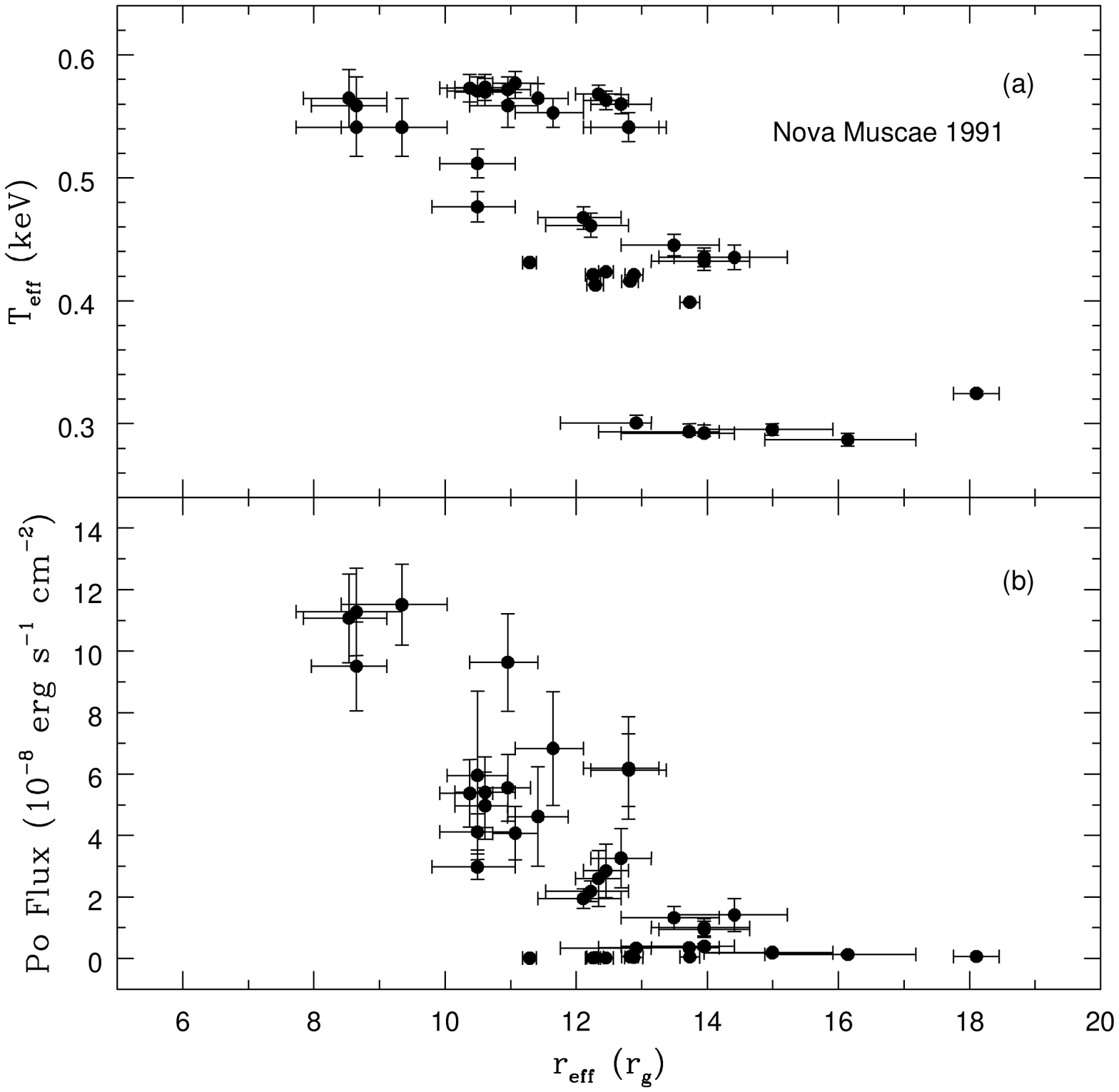}
\caption{ }
\end{figure}

\begin{figure}
\figurenum{9}
\plotone{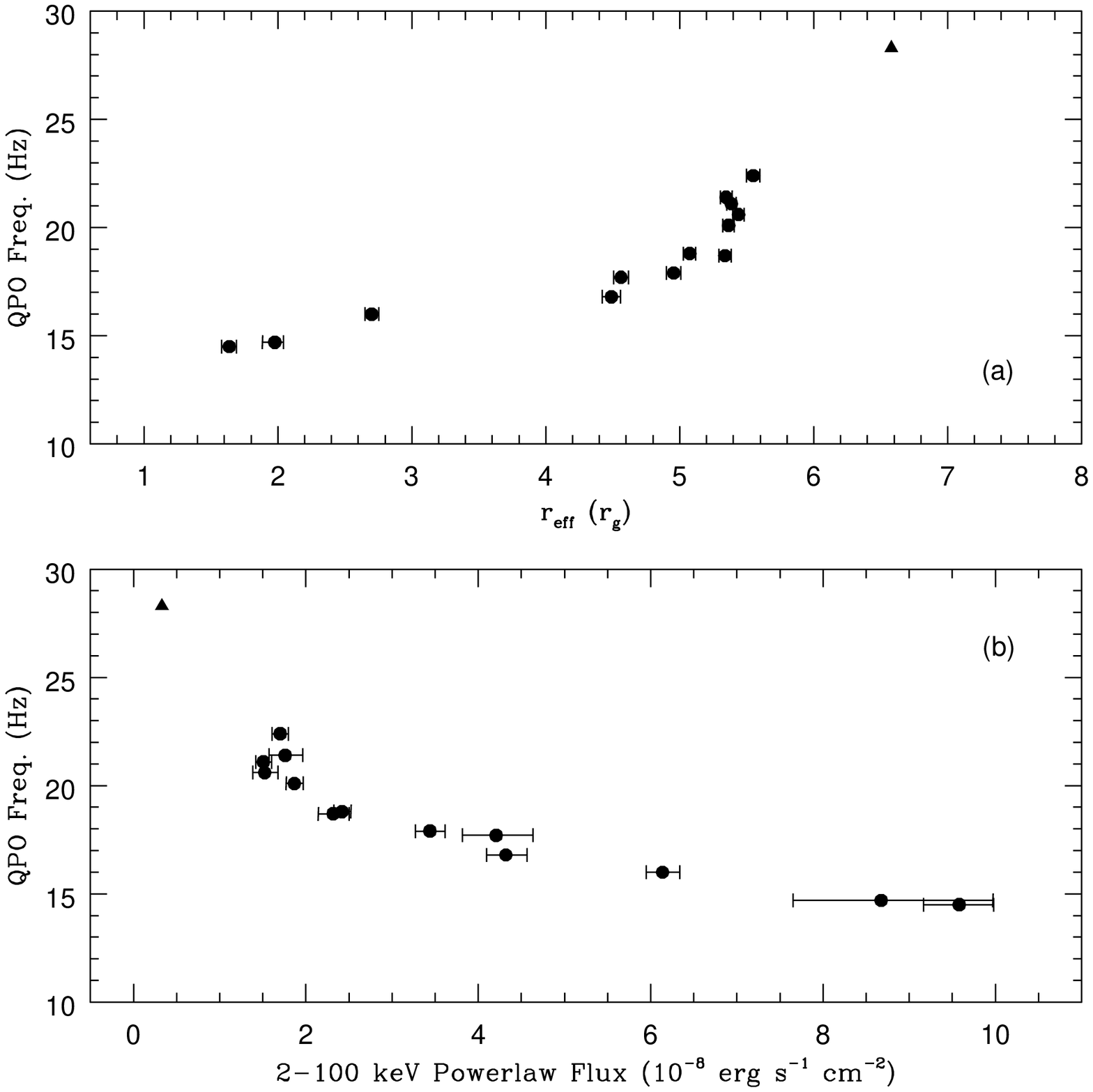}
\caption{ }
\end{figure}

\end{document}